\documentclass[showpacs,aps,prb,twocolumn,superscriptaddress]{revtex4} %\documentclass[showpacs,showkeys,aps,prb,preprint]{revtex4}

\usepackage[english]{babel} \usepackage{epsfig}
\usepackage{graphicx}
\usepackage{amsmath}
\usepackage{amssymb}
\usepackage{subfigure}
\usepackage{verbatim}
\usepackage{color}

\begin{document}

\title{Intrinsic arrested nanoscale phase separation near a topological Lifshitz transition in strongly correlated two-band metals}

\author{Antonio Bianconi} \affiliation{RICMASS - Rome International Center for Materials Science Superstripes, via dei Sabelli 119A, 00185 Rome, Italy}
\affiliation{Institute of Crystallography, CNR, via Salaria Km 29.300, 00015 Monterotondo, Rome, Italy}

\author{Nicola Poccia} \affiliation{RICMASS - Rome International Center for Materials Science Superstripes, via dei Sabelli 119A, 00185 Rome, Italy}
\affiliation{MESA + Institute for Nanotechnology, University of Twente, P.O. Box 217, 7500AE Enschede, Netherlands}

\author{A.~O.~Sboychakov} \affiliation{Institute for Theoretical and Applied Electrodynamics, Russian Academy of Sciences, Izhorskaya Str. 13, Moscow, 125412 Russia}

\author{A.~L.~Rakhmanov} \affiliation{Institute for Theoretical and Applied Electrodynamics, Russian Academy of Sciences, Izhorskaya Str. 13, Moscow, 125412 Russia}

\affiliation{Moscow Institute of Physics and Technology (State University), Institutskii Lane 9, Dolgoprudnyi, Moscow Region, 141700 Russia}

\affiliation{All-Russia Research Institute of Automatics, Moscow, 127055 Russia}

\author{K.~I.~Kugel} \affiliation{Institute for Theoretical and Applied Electrodynamics, Russian Academy of Sciences, Izhorskaya Str. 13, Moscow, 125412 Russia}

\begin{abstract}
The arrested nanoscale phase separation in a two-band Hubbard model for strongly correlated charge carriers is shown to occur in a particular range in vicinity of the topological Lifshitz transition, where the Fermi energy crosses the bottom of the narrow band and a new sheet of the Fermi surface related to the charge carriers of the second band comes into play. We determine the phase separation diagram of this two-band Hubbard model as a function of two variables, the charge carrier density and the energy shift between the chemical potential and the bottom of the second band. In this phase diagram, we first determine a line of quantum critical points for the Lifshitz transition and find criteria for the electronic phase separation resulting in an inhomogeneous charge distribution. Finally, we identify the critical point in presence of a variable long-range Coulomb interaction where the scale invariance of the coexisting phases with different charge densities appears. We argue that this point is relevant for the regime of scale invariance of the nanoscale phase separation in cuprates like it was first observed in La$_2$CuO$_{4.1}$.
\end{abstract}


\date{\today}

\maketitle

\section{Introduction}\label{intro}

The mechanism driving the emergence of a quantum macroscopic order that is able to resist to the decoherence effect of high temperatures remains a major topic of research in condensed matter. The realization of this macroscopic quantum phase in doped cuprates close to the Mott insulator regime has stimulated a large amount of investigations on the physics of strongly correlated metals. Most of theoretical papers treated models of a homogeneous system made of a single electronic band (or models of multiple hybridized bands reduced to a single effective band), with a large Hubbard repulsion. There is a growing agreement that the solution of the problem  of high-$T_c$ superconductivity requires the correct description of the normal state where spin, charge, orbital, and lattice degrees of freedoms compete, with the formation of nanoscale puddles of spin density wave stripes, puddles of charge density wave stripes, and/or puddles of ordered mobile oxygen interstitials.

A lot of researchers feel very strongly that the minimum model to capture the essential physics of high-temperature superconductors needs to take into account both the presence of ``two electronic components with different orbital symmetry"~\cite{00mc,0mc,1mc,2mc,3mc,4mc,5mc,6mc,7mc}, and a ``nanoscale phase separation"~\cite{1r,2r,3r,4r,5r,6r,7r,8r,9r} involving also the spatial segregation of the charge density, the orbital symmetry, and the lattice local symmetry~\cite{0nps,1nps,2nps,3nps,4nps,5nps,6nps,7nps,8nps,9nps,10nps,11nps,12nps}.
Therefore, a  multiband model is needed to describe the functional superconducting phase emerging in a complex system with multiple electronic components.~\cite{1mb,2mb,3mb,4mb,5mb,6mb} The effects of strong correlations in multiband systems were actively treated using the Hubbard model.~\cite{1mhm,2mhm,3mhm,4mhm,5mhm,6mhm,7mhm,8mhm} A particular interesting feature of the multiband Hubbard model is that it predicts the emergence of phase separation.~\cite{8mhm,Lorenzana,Main_PRL,Sboych_PRB2007}

In 1994 a topological Lifshitz transition~\cite{lifs0,lifs1,lifs2} was first proposed to appear around 1/8 doping in cuprates~\cite{1lif,2lif,3lif} and a theory for high-$T_c$ superconductivity based on the shape resonances between a BCS-like superconducting gap and a second gap in the BEC-BCS crossover regime in the new appearing band was formulated.~\cite{4lif,5lif,6lif,7lif,8lif} There is now compelling experimental evidence that the high temperature superconductivity emerges in the proximity to a topological Lifshitz transition.~\cite{9lif,10lif,11lif,12lif,13lif,14lif,15lif,16lif}

Here we provide a theoretical model for the phase diagram region where the nanoscale phase separation emergences in a two-band scenario of two strongly correlated electronic fluids in the proximity of a topological Lifshitz transition (so called 2.5 order transition). This simple model captures the key physics of the anomalous normal phase in cuprates exhibiting the phase separation as a function of charge density and the energy splitting between the two bands. This provides an additional insight into specific features of superconducting phases in different cuprate families, i.e., the new 3D phase diagram where the critical temperature depends on the doping and misfit strain between the active atomic layers and the spacer layers.~\cite{1strain,2strain,3strain,3nps}  There exists an evidence of two types of phase separation in cuprates (a) the phase separation in the underdoped regime, near the Mott phase, between a hole-poor antiferromagnetic phase and a metallic hole-rich phase and (b) the phase separation between two metallic phases, namely, between a hole-poor phase with doping close to 1/8 and a hole-rich phase with doping close to 1/4. The cuprates at optimum doping present the second type of phase separation as we have proposed before.~\cite{KuBian_PRB2008,KuBian_SST2009}  Recently it has been found that some cuprate systems like La$_2$CuO$_{4.1}$ show scale invariance of the distribution of oxygen interstitials that suggests a scale invariant phase separation typical of a system near the critical point. Therefore, it is possible that the criticality in La$_2$CuO$_{4+y}$ results from a quantum critical point.~\cite{6nps}
We discuss the phase diagram of a two-band system as a function of two variables: the charge density and the energy shift between the two bands. In this phase diagram, we first determine a line of quantum critical points for a Lifshitz transition of the type ``appearing of a spot" of a new sheet of the Fermi surface when one more band comes into play. Second, we identify the electronic phase separation for two strongly correlated bands in the proximity of the line of Lifshitz transition. Finally, we identify the critical point, where the phase invariance in the coexistence of the two phases appears. This last point is proposed to be a possible explanation for the regime of scale invariance in nanoscale phase separation in high-$T_c$ superconductors.~\cite{10nps}

\section{The model}\label{model}

\begin{figure}[t]
\centering
\includegraphics[width=0.95\columnwidth]{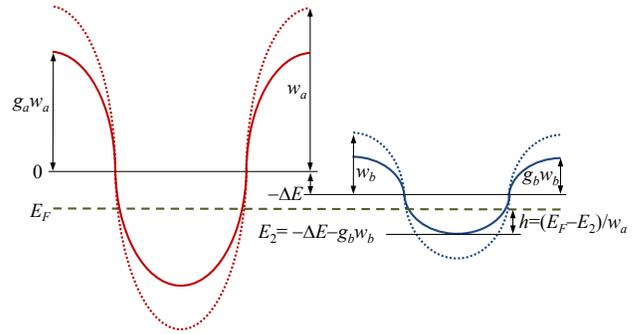}
\caption{Schematics of the band structure of Hamiltonian~\eqref{H}. There is a wide ($a$) and a narrow ($b$) correlated (lower Hubbard) bands shown by the solid cosine-like curves. The half-widths of these bands are $\bar{w}_{\alpha}=g_{\alpha}w_{\alpha}$ ($\alpha=a,b$), where $w_{\alpha}=zt_{\alpha}$ are half-widths of the bare (non-correlated) bands shown by the dotted cosine curves, and $g_{\alpha}$ are given by Eq.~\eqref{g}. The center of the wide band is chosen as zero energy. The center of the narrow band is shifted by the value $-\Delta E$. The Lifshitz parameter $h$ is defined as the position of the Fermi level $E_F$ relative to the bottom of the narrow band $E_2$ (in units of $w_a$).}\label{FigBands}
\end{figure}

The existence of the two types of the strongly correlated charge carriers in cuprates can be described in terms of the two-band Hubbard model. The Hamiltonian of such a system can be written as~\cite{Sboych_PRB2007}
\begin{eqnarray}\label{H}
H\!&=&\!-\!\!\!\sum_{\langle\mathbf{nm}\rangle\alpha,\sigma}\!\!t_{\alpha}a^
{\dag}_{\mathbf{n}\alpha\sigma}a_{\mathbf{m}\alpha\sigma}%
-\Delta E\sum_{\mathbf{n}\sigma}n_{\mathbf{n}b\sigma}
-E_F\sum_{\mathbf{n}\alpha,\sigma}n_{\mathbf{n}\alpha\sigma}\nonumber
\\
&+&\frac{1}{2}\sum_{\mathbf{n}\alpha,\sigma}U^{\alpha}n_{\mathbf{n}
\alpha\sigma}n_{\mathbf{n}\alpha\bar{\sigma}}%
+\frac{U'}{2}\sum_{\mathbf{n}\alpha,\sigma\sigma'}
n_{\mathbf{n}\alpha\sigma}n_{\mathbf{n}\bar{\alpha}\sigma'}\,.
\end{eqnarray}
Here, $a^{\dag}_{\mathbf{n}\alpha\sigma}$ and $a_{\mathbf{n}\alpha\sigma}$ are the creation and annihilation operators for electrons corresponding to bands $\alpha=\{a,\,b\}$ at site $\mathbf{n}$ with spin projection $\sigma$, and $n_{\mathbf{n}\alpha\sigma}=a^{\dag}_{\mathbf{n} \alpha\sigma}a_{\mathbf{n}\alpha\sigma}$. The symbol $\langle\dots\rangle$ denotes the summation over the nearest-neighbor sites. The first term in the right-hand side of Eq.~\eqref{H} corresponds to the kinetic energy of the conduction electrons in bands $a$ and $b$ with the hopping integrals $t_a>t_b$. In our model, we ignore the interband hopping. The second term describes the shift $\Delta E$ of the center of band $b$ with respect to the center of band $a$ ($\Delta E>0$ if the center of band $b$ is below the center of band $a$). The last two terms describe the on-site Coulomb repulsion of two electrons either in the same state (with the Coulomb energy $U^{\alpha}$) or in the different states ($U'$). The bar above $\alpha$ or $\sigma$ denotes {\it not} $\alpha$ or {\it not} $\sigma$, respectively. The assumption of the strong electron correlations means that the Coulomb interaction is large, that is, $U^{\alpha},\,U'\gg t_{\alpha},\,\Delta E$. The total number $n$ of electrons per site is a sum of electrons in the $a$ and $b$ states, $n=n_a+n_b$, and $E_F$ is the Fermi energy potential. Below, we consider the case $n\leq 1$ relevant to cuprates.  The model Eq.~\eqref{H} predicts a tendency to the phase separation in a certain range of parameters, in particular, in the case when the hopping integrals for $a$ and $b$ bands differ significantly ($t_a > t_b$)~\cite{Sboych_PRB2007,SboychDMFT}. This tendency results from the effect of strong correlations giving rise to dependence of the width of one band on the filling of another band. In the absence of the electron correlations ($n \ll 1$), the half-width $w_a = zt_a$ of $a$ band is larger than $w_b = zt_b$ ($z$ is the number of the nearest neighbors of the copper ion). Due to the electron correlations, the relative width of $a$ and $b$ bands can vary significantly~\cite{Sboych_PRB2007}. The schematic band structure and all notation are presented in Fig.~\ref{FigBands}.

%%%%%%%%%%%%%% Fig. 1 %%%%%%%%%%%%%%%

Following Ref.~\onlinecite{Sboych_PRB2007,KuBian_PRB2008} we considered the limit of strong correlations and introduce the one-particle Green's function,
\begin{equation}\label{GreenF}
G_{\alpha\sigma}(\mathbf{n} - \mathbf{n}_0,t - t_0) = - i\langle \hat{T}a_{\mathbf{n}\alpha\sigma}(t)a^\dag_{\mathbf{n}_0\alpha\sigma}(t_0)\rangle,
\end{equation}
where $\hat{T}$ is the time-ordering operator. The equations of motion for the one-particle Green's function with the Hamiltonian Eq.~\eqref{H} include the two-particle Green's functions. However, in the limit of strong on-site Coulomb repulsion, the presence of two electrons at the same site is unfavorable, and the two-particle Green's function is of the order of $1/U$, where $U\sim U_\alpha,U'$. In turn, the equation of motion for the two-particle Green's functions includes the three-particle terms, which are of the order of $1/U^2$ and so on. We use for the two-particle Green's functions the Hubbard I approximation and neglect the terms of the order of $1/U^2$. In so doing, we get a closed system for the one- and two-particle Green's functions~\cite{Sboych_PRB2007,KuBian_PRB2008}. This system is solved in a standard manner by passing from the space--time $(\mathbf{r},t)$ to the momentum--frequency $(\mathbf{k},\omega)$ representation. In the case of superconductors the number of electrons per site $n\leq1$. The upper Hubbard sub-bands are empty, and we can proceed to the limit $U^\alpha,U'\rightarrow\infty$.  In this case, the one-particle Green's function is independent of $U$ and can be written in the form~\cite{Sboych_PRB2007,KuBian_PRB2008}
\begin{equation}\label{GreenKO}
G_{\alpha\sigma}(\mathbf{k},\omega) = \frac{g_{\alpha\sigma}}{\omega+E_F+\Delta E^\alpha-g_{\alpha\sigma}w_\alpha\zeta(\mathbf{k})},
\end{equation}
where $\Delta E^a =0$, $\Delta E^b =\Delta E$,
\begin{equation}\label{g}
g_{\alpha\sigma}=1-\sum_{\sigma'}n_{\bar{\alpha}\sigma'}-n_{\alpha\bar{\sigma}},
\end{equation}
$n_{\alpha\sigma}=\langle n_{\mathbf{n}\alpha\sigma}\rangle$ is the average number of electrons per site in the state $(\alpha,\sigma)$, and $\zeta(\mathbf{k})$ is the spectral function depending on the lattice symmetry. In the main approximation in $1/U$, the magnetic ordering does not appear and we can assume that $n_{\alpha\uparrow}=n_{\alpha\downarrow}=n_\alpha/2$ and $g_{\alpha\uparrow}=g_{\alpha\downarrow}\equiv g_\alpha$. For simplicity and for more direct comparison with the results of Ref.~\onlinecite{Sboych_PRB2007}, we use here the dispersion law corresponding to the tight-binding band in the simple cubic lattice, $\zeta(\mathbf{k})=-\left[\cos{(k^1d)}+\cos{(k^2d)}\cos{(k^3d)}\right]/3$, where $d$ is the lattice parameter. We checked that the qualitative results do not significantly affected by the specific choice of the dispersion law. However, for a more detailed comparison of the model predictions with  the actual experimental data, it is necessary to use realistic electronic characteristics. This work is now in progress.

It follows from Eqs.~\eqref{GreenKO} and ~\eqref{g} that the filling of band $a$ depends on the filling of band $b$ and {\it vice versa}. Really, using the expression for the density of states $\rho_\alpha(E)=-\pi^{-1}\textrm{Im}\int{G_\alpha(\mathbf{k},E+i0)d^3\mathbf{k}/(2\pi)^3}$, we get the expression for the numbers of electrons in bands $a$ and $b$
\begin{equation}\label{nn}
n_\alpha=2g_\alpha n_0\left(\frac{E_F+\Delta E^\alpha}{g_\alpha w_\alpha}\right),
\end{equation}
where~\cite{Sboych_PRB2007,KuBian_PRB2008}
\begin{equation}\label{n0}
n_0(\mu')=\int_{-1}^{\mu'}{dE'\rho_0(E')},
\end{equation}
and $\rho_0(E')=\int{\delta[E'-\zeta(\mathbf{k})]d^3\mathbf{k}/(2\pi)^3}$ is the density of states for free electrons. The Fermi level, $E_F$, in Eq.~\eqref{nn} is found from the equality $n=n_a(E_F)+n_b(E_F)$.

In iron-based superconductors, as it was shown in Refs.~\onlinecite{6lif,7lif,Fernandes_PRB2010}, the region of high $T_c$ appears in the neighborhood of the Lifshitz transition where the local Fermi surface spot disappears. The Lifshitz transition is a common feature of many types of superconductors and in its neighborhood the standard BCS approach is hardly applicable. The situation here bears a similarity with the BEC--BCS crossover widely studied in the physics of ultracold atomic systems.  In the specific case of strongly correlated electron systems including two bands (two types of charge carriers), the shift of the chemical potential due to the relative shift of the bands and/or the variation of charge density implies the relevant renormalization  of the effective width of both bands. This strongly nonlinear renormalization leads to the electronic phase separation. Since in the high-$T_c$ superconductors an increase of the critical temperature occurs at a substantiable distance from the Lifshitz transition, it is tempting to associate  the region of the phase separation with that corresponding to high values of the critical temperature. The experimental evidence suggests that the phase separation goes together with the high-$T_c$ superconductivity. In this paper, we calculate the region of the phase separation as a function of the Lifshitz parameter.

\section{Results}\label{results}

%%%%%%%%%%%%%% Fig. 2 %%%%%%%%%%%%%%%
\begin{figure}     \centering
\subfigure[$\quad \delta = 0.125$]
{\includegraphics[width=0.9\columnwidth]{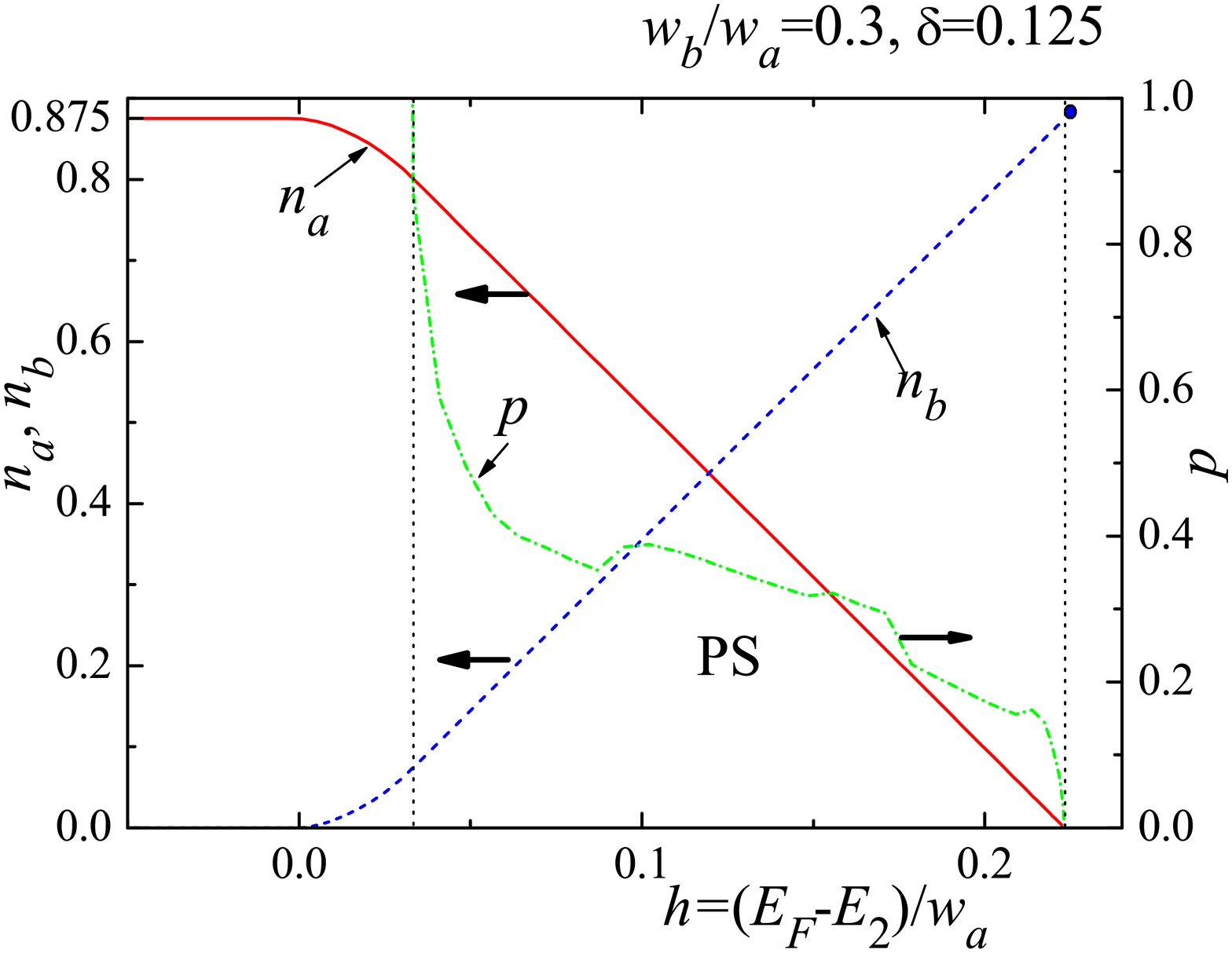} } \subfigure[$\quad \delta = 0.3$]
{\includegraphics[width=0.9\columnwidth]{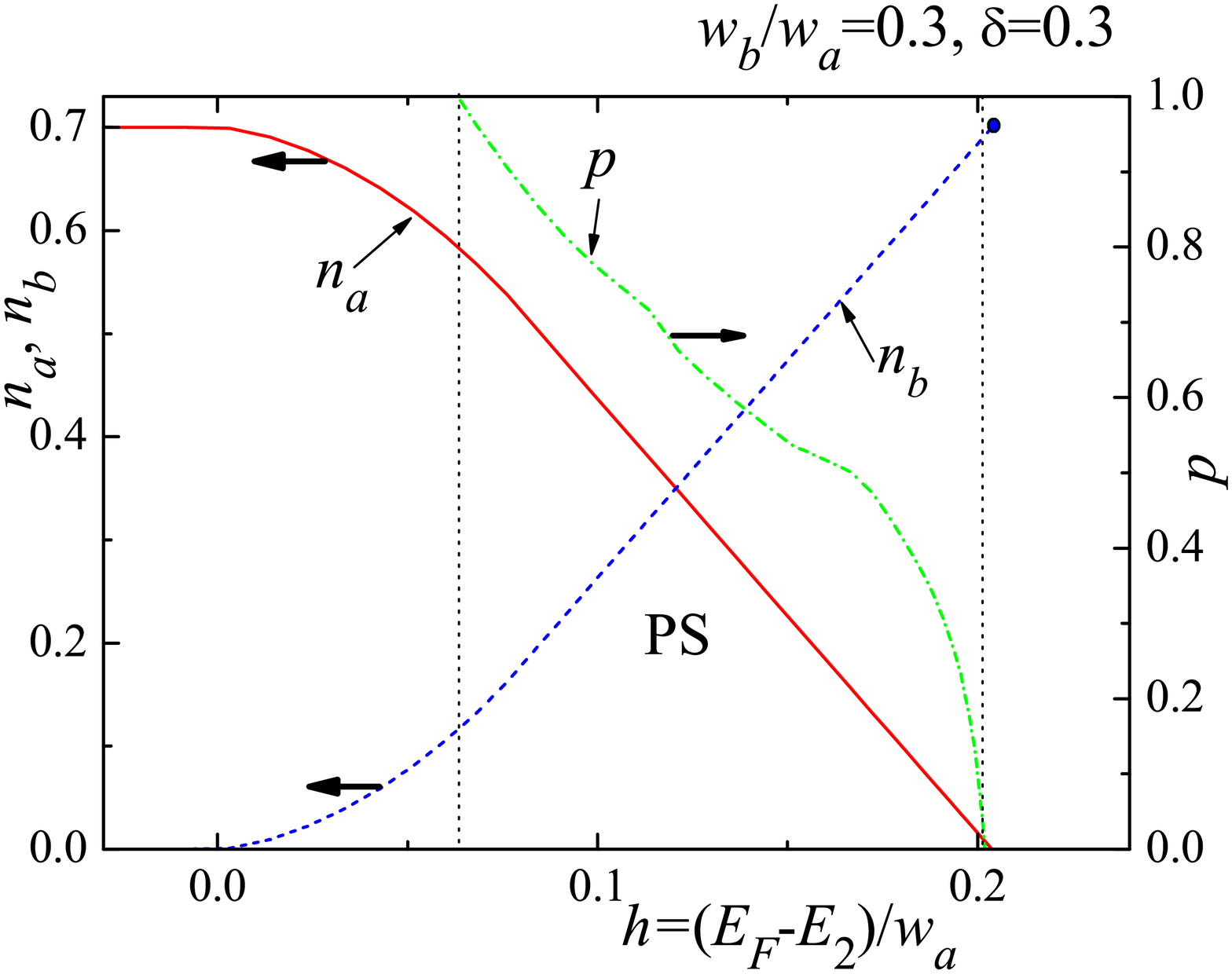} } \subfigure[$\quad \delta = 0.5$]
{\includegraphics[width=0.9\columnwidth]{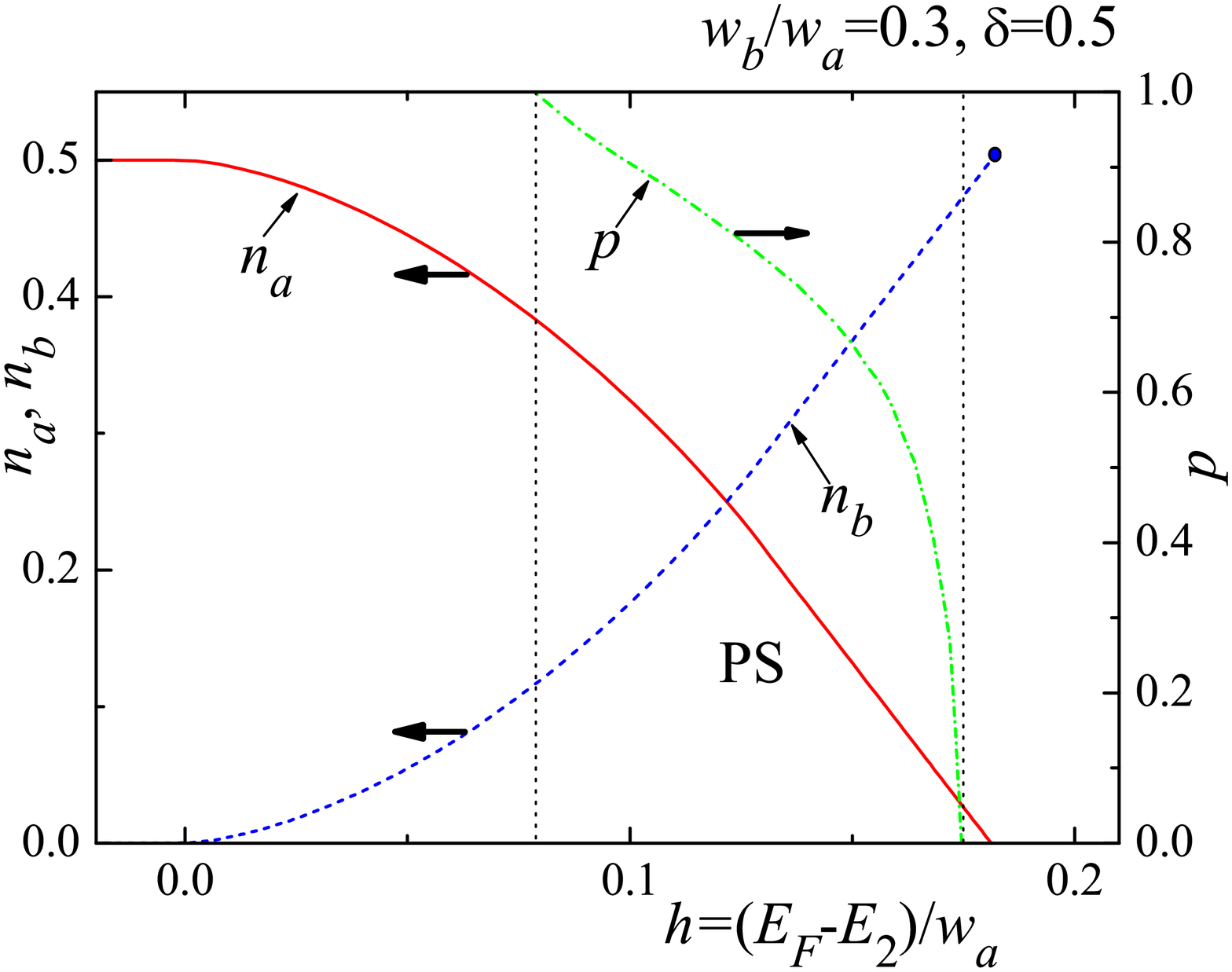} } \caption{Evolution of the occupation numbers $n_a$ and $n_b$ of the bands $a$ and $b$ at different doping levels $\delta =1-n$ in the absence of phase separation. The region of phase separation lies between two vertical dotted lines. There we have two phases: $P_a$, including mostly $a$ charge carriers and $P_b$ with dominant $b$ carriers. The content of different types of carriers in $P_a$ and $P_b$ is given by the intersections of $n_a$ and $n_b$ curves with left and right dashed vertical curves, respectively. The change in the volume fraction $p$ of phase $P_a$ in the phase separation region is shown by the (green) dot-dashed line.
An irregular shape of the $p(h)$ (green) curve in panel (a) is due to small difference between the energies of the homogeneous and the phase separated state at low doping.} \label{Fig1}
\end{figure}

Poles of the Green's function Eq.~\eqref{GreenKO} give two energy bands of our model. The Lifshitz parameter $h=(E_F-E_2)/w_a$ determines how far is the position of the Fermi level $E_F$ from the bottom $E_2$ of the narrow band $b$ (see Fig.~\ref{FigBands}). For $h>0$ the charge carriers of the $b$ type exist in the system. At fixed doping level $\delta=1-n=1-n_a-n_b$, the occupation numbers, $n_a$ and $n_b$ depend on the value of $h$. The dependence of the filling of bands $a$ and $b$ on the Lifshitz parameter is non-trivial for strongly correlated bands because the widths of these bands, in turn, depend on the fillings $n_a$ and $n_b$. We calculate the dependence of $n_a$ and $n_b$ on the Lifshitz parameter $h$ according to the approach developed in Refs.~\onlinecite{Main_PRL,Sboych_PRB2007}. The obtained curves for three different doping levels $\delta$ are shown in Fig.~\ref{Fig1}. These dependences are qualitatively similar. Electrons appear in band $b$ if $h>0$. Simultaneously, the number of electrons in the $a$ band starts to decrease and it goes to zero at some critical value of the Lifshitz parameter.

%%%%%%%%%%%%%%%%%% Fig. 3 %%%%%%%%%%%%%%%%%%%
\begin{figure}\centering \includegraphics[width=0.9\columnwidth]{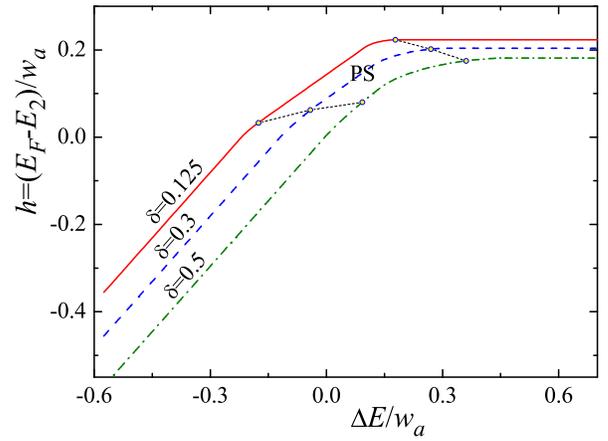} \caption{(Color online) Distance $h$ between the Fermi level and the bottom of the upper narrow band (Lifshitz parameter) versus the shift $\Delta E$ between two bands at different doping $\delta = 1- n$.} \label{Fig2} \end{figure}

We postulated that the ground state of the system is homogeneous when obtaining the above results. The analysis performed in Refs.~\onlinecite{Main_PRL,Sboych_PRB2007} shows, however, that this is not so in general case. Indeed, the energy of the system in the homogeneous state, $E_\textrm{hom}$, is the sum of electron energies in all filled bands. We can write $E_\textrm{hom}$ in the form~\cite{KuBian_PRB2008}
\begin{eqnarray}\label{Ehom}
\nonumber
  E_\textrm{hom} &=& 2\sum_\alpha g^2_\alpha w_\alpha\varepsilon_0\left(\frac{E_F+\Delta E^\alpha}{g_\alpha w_\alpha}\right) ,\\
  \varepsilon_0(\mu') &=& \int_{-1}^{\mu'}{dE'E'\rho_0(E')}.
\end{eqnarray}
The analysis of these equations reveals that within a certain $n$ range the system compressibility is negative, $\partial^2 E_{\textrm{hom}}\partial n^2<0$,~\cite{KuBian_PRB2008} which means a possibility for the charge carriers to form two phases with different electron concentrations.

The electronic phase separation occurs in a wide range of model parameters and doping levels. At fixed doping, the phase-separated state is the ground state of the system if the Lifshitz parameter lies within definite limits $h_1<h<h_2$ (see vertical lines in Figs.~\ref{Fig1}a-c). The separated phases are $P_a$ with total ($a$ and $b$) electron concentration $n_1$, and $P_b$ having a different electron concentration $n_2$. For the phase $a$ ($b$) the electrons of $a$ ($b$) type are dominant, that is, $n_a\gg n_b$ ($n_b\gg n_a$). The volume fraction $p$ of the phase $P_a$, as well as concentrations $n_1$ and $n_2$, can be found by the minimization of the system's energy, $E_{ps}=pE_{\textrm{hom}}(n_1)+(1-p)E_{\textrm{hom}}(n_2)$ with the condition $1-\delta =pn_1+(1-p)n_2$. The value of $p$ decreases from $p=1$ down to zero for $h$ changing from $h_1$ to $h_2$ as shown in Fig.~\ref{Fig1}.

%%%%%%%%%%%%%%%%%%% Fig. 4 %%%%%%%%%%%%%%%%%%
\begin{figure} \centering \includegraphics[width=0.9\columnwidth]{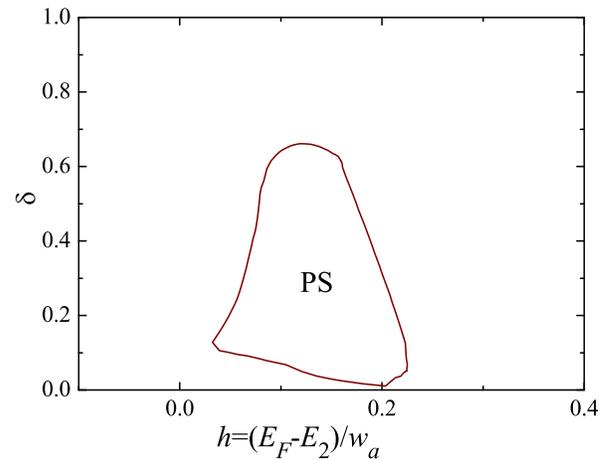} \caption{(Color online) The phase separation region for the ratio of the band widths $w_b/w_a = 0.3$.} \label{Fig3} \end{figure}

The Lifshitz parameter depends both on the doping $\delta$ (via the position of the Fermi level) and the energy shift between the centers of two bands $\Delta E$. At fixed doping level, there is one-to-one correspondence between $\Delta E$ and $h$. Typical curves $h\left(\Delta E/w_a\right)$ are shown in Fig.~\ref{Fig2} for different $\delta$. The phase separation exists in the region restricted by two black dotted curves. In Ref.~\onlinecite{Sboych_PRB2007}, the phase diagram of the two-band Hubbard model~\eqref{H} in the plane ($n$,$\Delta E$) has been obtained in the limit of large $U$. Using these results and the relation between $h$ and $\Delta E$ for different doping levels, we can rebuild this phase diagram in the plane ($h$,$\delta$). The result is shown in Fig.~\ref{Fig3}. The phase separation exists within the region restricted by the (red) solid contour.

%%%%%%%%%%%%%%%% Fig. 5 %%%%%%%%%%%%%%%%%%%%%
\begin{figure}
\centering \includegraphics[width=0.9\columnwidth]{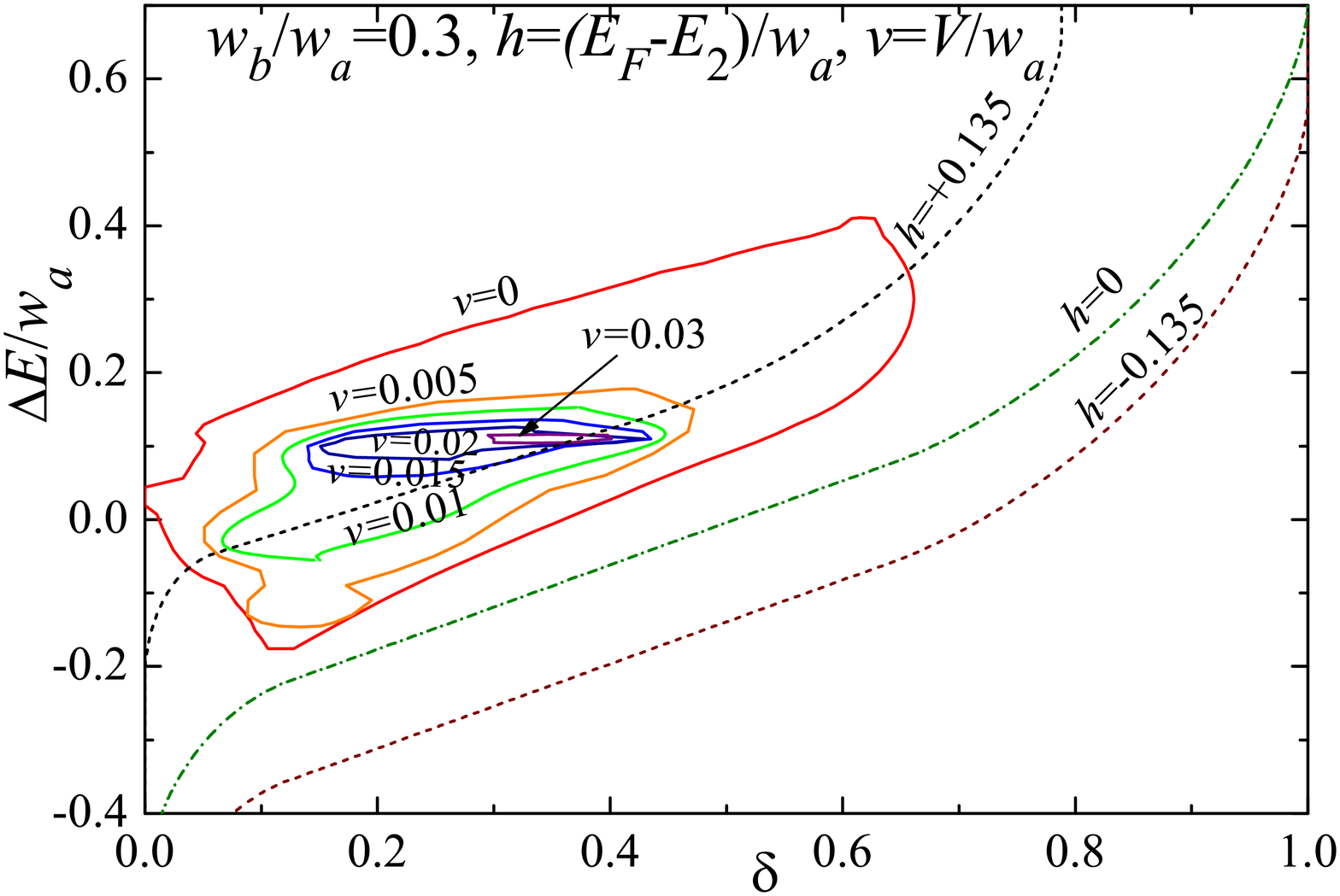}
\caption{The phase separation region in the phase diagram of model Eq.~\eqref{H} at the ratio of band widths $w_b/w_a = 0.3$.  The charge neutrality breaking in the phase separated state substantially reduces this region. It shrinks with the growth of  $v = V/w_a$, that is, with the growth of the long-range Coulomb interaction.} \label{Fig4}
\end{figure}

The phase separation discussed above gives rise to the breaking of the local charge neutrality since the charge carrier concentration is different in different phases. Thus, we should take into account an additional electrostatic contribution in the free energy, $E_C$, which is governed by the long-range Coulomb interaction (this contribution has been neglected in the above discussion). This term in the Wigner--Seitz approximation was calculated in Refs.~\onlinecite{Sboych_PRB2007,KuBian_PRB2008,DiCastroPS}. If $p<0.5$, it can be written as $E_C=V(n_1-n_2)^2(R/d)^2u(p)$, where $u(p)=2\pi p(2-2p^{1/3}+p)/5$, $V=e^2/\varepsilon d$ is the characteristic energy of the intersite Coulomb interaction, $e$ is the elementary charge, $\varepsilon$ is the long-range permittivity, and $R$ is the radius of the spherical droplet of the phase $P_a$ surrounded by the shell of the phase $P_b$. In the case $p>0.5$, we should replace $n_1 \leftrightarrow n_2$ and $p\leftrightarrow 1-p$.

The value of $E_C$ decreases with decreasing a spatial scale of the inhomogeneous state. However, the smaller is the characteristic size of inhomogeneity, the higher is the energy of the phase interface $E_S$. We assumed above that the phase with lower volume fraction $p$ forms spheres of the radius $R$ located in the matrix of another phase. In this case, the energy of the phase interface $E_S$ can be written as $E_S=3p\sigma d/R$, where $\sigma$ is the interface tension, which we calculate using the Balian--Bloch perturbative approach~\cite{BalianBloch}. Such calculations are described in detail in Ref.~\onlinecite{Sboych_PRB2007}. Minimizing $E_{CS}=E_C+E_S$ with respect to $R$, we obtain the characteristic scale of the phase-separated state and get more realistic estimate for the free energy of the inhomogeneous system~\cite{Lorenzana,DiCastroPS}. The optimized value of $E_{CS}$ is given by the following relation~\cite{Sboych_PRB2007}:
\begin{equation}
E_{CS}=\frac32\left[18p^2\sigma^2(n_1-n_2)^2u(p)V\right]^{1/3}\,.
\end{equation}
As follows from this formula, the new contribution to the total free energy depends on the long-range Coulomb repulsion parameter as $E_{CS}\propto V^{1/3}$. The region of parameters, where the phase separation is favorable, shrinks with the increase of $E_{CS}$, that is, with the growth of the long-range Coulomb repulsion $V$ and disappears if this value is above some threshold.

In other words, the long-range Coulomb interaction induces a shrinkage of the phase separation region together with the scale of  the phase separation. Hence we can say that here we deal with the frustrated (or arrested) phase separation. Note that the term ``frustrated phase separation" was first introduced by Emery and Kivelson~\cite{1frPS} for strongly correlated electron systems and is rather widely used in this field (see, e.g. Refs.~\onlinecite{Lorenzana,2frPS,3frPS}), whereas the synonym of this term, namely, ``arrested phase separation"  has been used long  before but mainly in relation to colloidal solutions and gels (see, e.g. Refs.~\onlinecite{1arPS,2arPS,3arPS}) and now it is used in a more general context.~\cite{4arPS,5arPS} We believe that the word ``arrested" is more adequate here and prefer to use it.

The phase separation region is shown in Fig.~\ref{Fig4} in the plane $(\delta,\Delta E/w_a)$ for different values of $v=V/w_a$. The long-range Coulomb repulsion affects significantly the phase separation region (if $v>10^{-3}$ for the chosen range of parameters). The area of the inhomogeneous state rapidly shrinks (if $v>0.005$ in Fig.~\ref{Fig4}) and totally disappears if $v>v_c$ ($v_c\approx 0.03$ in Fig.~\ref{Fig4}). The values $V/w$ in Fig.~\ref{Fig4} are realistic for high-$T_c$ cuprates.~\cite{KuBian_SST2009}

%%%%%%%%%%%%%%%%%%%%%% Fig. 6 %%%%%%%%%%%%%%%%
\begin{figure}\centering \includegraphics[width=0.95\columnwidth]{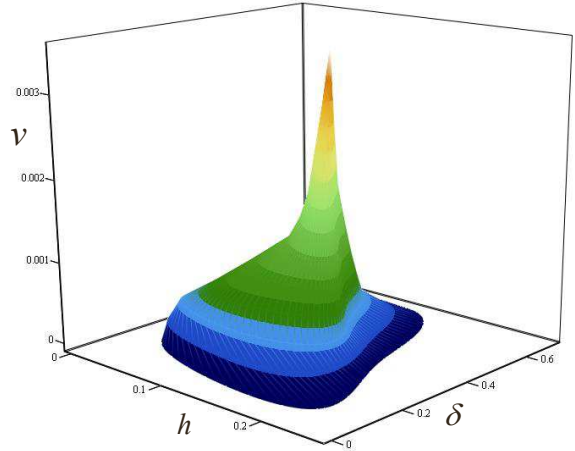}\caption{The three-dimensional phase diagram of model \eqref{H} generalizing the data presented in the previous figures.} \label{Fig3D} \end{figure}

The phase separation in the two-band model is possible only in the vicinity of the Lifshitz transition, that is, in definite range of parameter $h$. In Fig.~\ref{Fig4} the lines of constant $h$ are shown by dotted lines. The phase separation is evidently possible only if $h>0$. In Fig.~\ref{Fig3D}, the region of the phase separation is shown in three-dimensional phase diagram in the space ($h$,$\delta$,$v$). This figure summarizes the results of our calculations. The inhomogeneous state exists in a definite region of doping and Lifshitz parameter. This region decreases with the increase of the long-range Coulomb repulsion parameter $v=V/w_a$ and shrinks to zero if $v>v_c$. We can say that the shrinkage of the phase-separation region allow the charge carrier densities in the  phase-separated state to be closer to the line of Lifshitz transition.

\section{Discussion}\label{discus}

Now the point is where high $T_c$ occurs in a two-band scenario. The detailed discussion of this issue is given in Refs.~\onlinecite{6lif} and \onlinecite{7lif}.
Let us move the bottom $E_2$ of the second band relative to the Fermi level and we shall deal with the following two regimes.

1. The system is boson-fermion regime with a low $T_c$, where a first ``BCS condensate" resonates with a ``BEC condensate", for the negative Lifshitz parameter, $-w_0/w_a<h<0$, where $h = (E_F -E_2)/w_a$, $w_0$ is the cutoff energy for the pairing interaction, and $w_a$ is the width of the first band.

2. At the ``shape resonance" in an optimum regime, where a first ``BCS condensate" in an electron-rich band resonates with a second ``condensate at the BEC-BCS crossover" occurring for a  positive values of $h$, the critical temperature starts increasing and attains maximum at $h$ of the order of $w_0/w_a$.

Now the problem is that in this range of the tuning of the chemical potential, the phase separation also occurs. Moreover, in oxygen doped system we have identified, where the critical point for phase separation appears and it is quite near to the $w_0/w_a = h$ range.  Therefore the distance in energy ($h$ in our notation) of the critical point from the band edge could be a measure of the unknown energy cutoff for the pairing interaction in cuprates. These ideas are illustrated by the figures presented in the previous section.  The undoped state of the cuprates corresponds to one electron per site ($n = 1$) in the model used in Ref.~\onlinecite{Sboych_PRB2007}. The number of itinerant holes $\delta$ is related to $n$ as $\delta = 1 - n$. In general, the relationship between $n$ and $\delta$ could be more complicated~\cite{5mc}, however, for the present considerations such corrections are not of principal importance.

In conclusion, we can say that our simplified model provides a good illustration for general ideas that high-$T_c$ superconductivity is an inherent feature of functional ``heterostructures at atomic limit" made of atomic units, where four essential ingredients are well tuned. (1) Two or more electronic components give multiple Fermi surface spots with different symmetry so that (a) single electron interband hopping is forbidden while (b) interband exchange-like pair transfer is allowed. (2) The Fermi energy of one of the components is close to the band edge so the system is close to the 2.5 order Lifshitz (metal-to-metal) transition. (3) The lattice and electronic structure show the complex granular ``superstripes" matter: a nanoscale phase separation made of superconducting puddles coexisting with normal stripes with charge order (CDW) and/or magnetic puddles with spin order (SDW), which does not suppress but enhances the stability of the macroscopic quantum order. (4) Intragrain high-$T_c$ superconductivity is controlled by the ``shape resonances" between a first BCS condensate and a second condensate in the BEC-BCS crossover. Therefore, further essential details are needed to investigate in the scenario of multi-condensates superconductivity in the regime of percolation superconductivity corresponding to establishing the long-range coherence in scale-free networks.~\cite{4r,5r}

In this work, we have shown that the synthesis of a two-band strongly correlated ``multi-condensate superconductor", where a first BCS condensate in a large Fermi surface coexists with a second  condensate at the BEC-BCS crossover in a new appearing small Fermi surface (like in cuprates and iron-based superconductors)~\cite{2lif, 3lif, 4lif, 5lif, 6lif, 7lif, 8lif, 9lif} should also exhibit an intrinsic arrested nanoscale phase separation. In fact, this type of complex superconductivity appears in a two-band metal at a critical distance from the topological Lifshitz transition. Moreover, the control of long-range Coulomb interaction~\cite{KuBian_PRB2008,KuBian_SST2009}, determined by the screening in the different materials surrounding metal units, is a needed key parameter to bring the system to a self-similar phase~\cite{10nps,11nps,12nps}, which will also promote~\cite{4r,5r} the high-$T_c$ superconductivity.

\section*{Acknowledgments}

The work was supported by Superstripes Institute, Dutch FOM and NWO foundations, and Russian Foundation for Basic Research, project Nos. 12-02-00339 and 14-02-00276. N.P. acknowledges support from the Marie Curie Intra-European Fellowship.

\end{document}